




\font\twelverm=cmr10 scaled 1200    \font\twelvei=cmmi10 scaled 1200
\font\twelvesy=cmsy10 scaled 1200   \font\twelveex=cmex10 scaled 1200
\font\twelvebf=cmbx10 scaled 1200   \font\twelvesl=cmsl10 scaled 1200
\font\twelvett=cmtt10 scaled 1200   \font\twelveit=cmti10 scaled 1200

\skewchar\twelvei='177   \skewchar\twelvesy='60


\def\twelvepoint{\normalbaselineskip=12.4pt
  \abovedisplayskip 12.4pt plus 3pt minus 9pt
  \belowdisplayskip 12.4pt plus 3pt minus 9pt
  \abovedisplayshortskip 0pt plus 3pt
  \belowdisplayshortskip 7.2pt plus 3pt minus 4pt
  \smallskipamount=3.6pt plus1.2pt minus1.2pt
  \medskipamount=7.2pt plus2.4pt minus2.4pt
  \bigskipamount=14.4pt plus4.8pt minus4.8pt
  \def\rm{\fam0\twelverm}          \def\it{\fam\itfam\twelveit}%
  \def\sl{\fam\slfam\twelvesl}     \def\bf{\fam\bffam\twelvebf}%
  \def\mit{\fam 1}                 \def\cal{\fam 2}%
  \def\tt{\twelvett}
  \textfont0=\twelverm   \scriptfont0=\tenrm   \scriptscriptfont0=\sevenrm
  \textfont1=\twelvei    \scriptfont1=\teni    \scriptscriptfont1=\seveni
  \textfont2=\twelvesy   \scriptfont2=\tensy   \scriptscriptfont2=\sevensy
  \textfont3=\twelveex   \scriptfont3=\twelveex  \scriptscriptfont3=\twelveex
  \textfont\itfam=\twelveit
  \textfont\slfam=\twelvesl
  \textfont\bffam=\twelvebf \scriptfont\bffam=\tenbf
  \scriptscriptfont\bffam=\sevenbf
  \normalbaselines\rm}



\def\beginlinemode{\endmode
  \begingroup\parskip=0pt \obeylines\def\\{\par}\def\endmode{\par\endgroup}}
\def\beginparmode{\endmode
  \begingroup \def\endmode{\par\endgroup}}
\let\endmode=\par
{\obeylines\gdef\
{}}
\def\singlespace{\baselineskip=\normalbaselineskip}

\def\oneandahalfspace{\baselineskip=\normalbaselineskip
  \multiply\baselineskip by 3 \divide\baselineskip by 2}
\def\doublespace{\baselineskip=\normalbaselineskip \multiply\baselineskip by 2}

\newcount\firstpageno
\firstpageno=2
\footline={\ifnum\pageno<\firstpageno{\hfil}\else{\hfil\twelverm\folio\hfil}\fi}
\let\rawfootnote=\footnote              
\def\footnote#1#2{{\rm\singlespace\parindent=0pt\rawfootnote{#1}{#2}}}
\def\raggedcenter{\leftskip=2em plus 12em \rightskip=\leftskip
  \parindent=0pt \parfillskip=0pt \spaceskip=.3333em \xspaceskip=.5em
  \pretolerance=9999 \tolerance=9999
  \hyphenpenalty=9999 \exhyphenpenalty=9999 }
\def\dateline{\rightline{\ifcase\month\or
  January\or February\or March\or April\or May\or June\or
  July\or August\or September\or October\or November\or December\fi
  \space\number\year}}
\def\received{\vskip 3pt plus 0.2fill
 \centerline{\sl (Received\space\ifcase\month\or
  January\or February\or March\or April\or May\or June\or
  July\or August\or September\or October\or November\or December\fi
  \qquad, \number\year)}}

\parskip=\medskipamount
\twelvepoint            
\doublespace            
\overfullrule=0pt       


\def\preprintno#1{
 \rightline{\rm #1}}    

\def\author                     
  {\vskip 3pt plus 0.2fill \beginlinemode
   \singlespace \raggedcenter}

\def\affil                      
  {\vskip 3pt plus 0.1fill \beginlinemode
   \oneandahalfspace \raggedcenter \sl}

\def\abstract                   
  {\vskip 3pt plus 0.3fill \beginparmode
   \doublespace \narrower ABSTRACT: }

\def\endtitlepage               
  {\endpage                     
   \body}

\def\body                       
  {\beginparmode}               

\def\head#1{                    
  \filbreak\vskip 0.5truein     
  {\immediate\write16{#1}
   \raggedcenter \uppercase{#1}\par}
   \nobreak\vskip 0.25truein\nobreak}

\def\subhead#1{                 
  \vskip 0.25truein             
  {\raggedcenter #1 \par}
   \nobreak\vskip 0.25truein\nobreak}

\def\refto#1{$|{#1}$}           

\def\references                 
  {\subhead{References}         
   \beginparmode
   \frenchspacing \parindent=0pt \leftskip=1truecm
   \parskip=8pt plus 3pt \everypar{\hangindent=\parindent}}

\gdef\refis#1{\indent\hbox to 0pt{\hss#1.~}}    

\gdef\journal#1, #2, #3, 1#4#5#6{               
    {\sl #1~}{\bf #2}, #3, (1#4#5#6)}           

\def\refstylenp{                
  \gdef\refto##1{ [##1]}                                
  \gdef\refis##1{\indent\hbox to 0pt{\hss##1)~}}        
  \gdef\journal##1, ##2, ##3, ##4 {                     
     {\sl ##1~}{\bf ##2~}(##3) ##4 }}

\def\refstyleprnp{              
  \gdef\refto##1{ [##1]}                                
  \gdef\refis##1{\indent\hbox to 0pt{\hss##1)~}}        
  \gdef\journal##1, ##2, ##3, 1##4##5##6{               
    {\sl ##1~}{\bf ##2~}(1##4##5##6) ##3}}

\def\endreferences{\body}

\def\figurecaptions             
  {\endpage
   \beginparmode
   \head{Figure Captions}
}

\def\endpage                    
  {\vfill\eject}

\def\endpaper                   
  {\endmode\vfill\supereject}

\def\endit
  {\endpaper\end}


\def\ref#1{Ref. #1}                     
\def\Ref#1{Ref. #1}                     

\def\frac#1#2{{\textstyle{#1 \over #2}}}

\def\m@th{\mathsurround=0pt }
\def\leftrightarrowfill{$\m@th \mathord\leftarrow \mkern-6mu
 \cleaders\hbox{$\mkern-2mu \mathord- \mkern-2mu$}\hfill
 \mkern-6mu \mathord\rightarrow$}
\def\overleftrightarrow#1{\vbox{\ialign{##\crcr
     \leftrightarrowfill\crcr\noalign{\kern-1pt\nointerlineskip}
     $\hfil\displaystyle{#1}\hfil$\crcr}}}


\font\titlefont=cmr10 scaled\magstep3

\def\oneandfourfifthsspace{\baselineskip=\normalbaselineskip
  \multiply\baselineskip by 9 \divide\baselineskip by 5}

\def\newstyletitle                      
  {\null\vskip 3pt plus 0.2fill
   \beginlinemode \doublespace \raggedcenter \titlefont}
\def\uof{Department of Physics\\University of Florida\\
Gainesville, FL 32611}

\def\heading                            
  {\vskip 0.5truein plus 0.1truein      
   \beginparmode \def\\{\par} \parskip=0pt \singlespace \raggedcenter}

\def\subheading                         
  {\vskip 0.25truein plus 0.1truein     
   \beginlinemode \singlespace \parskip=0pt \def\\{\par}\raggedcenter}

\def\tag#1$${\eqno(#1)$$}

\def\align#1$${\eqalign{#1}$$}

\def\aligntag#1$${\gdef\tag##1\\{&(##1)\cr}\eqalignno{#1\\}$$
  \gdef\tag##1$${\eqno(##1)$$}}

\def\endaligntag{}

\def\overset #1\to#2{{\mathop{#2}\limits^{#1}}}
\def\underset#1\to#2{{\let\next=#1\mathpalette\undersetpalette#2}}
\def\undersetpalette#1#2{\vtop{\baselineskip0pt
\ialign{$\mathsurround=0pt #1\hfil##\hfil$\crcr#2\crcr\next\crcr}}}


\def\[#1]{[\cite{#1}]}
\def\cite#1{{#1}}
\def\(#1){(\call{#1})}
\def\call#1{{#1}}
\def\taghead#1{}
\def\frac#1#2{{#1 \over #2}}

\def\sla{\raise.15ex\hbox{$/$}\kern-.57em}
\def\leaderfill{\leaders\hbox to 1em{\hss.\hss}\hfill}
\def\twiddle{\lower.9ex\rlap{$\kern-.1em\scriptstyle\sim$}}
\def\bigtwiddle{\lower1.ex\rlap{$\sim$}}
\def\gtwid{\mathrel{\raise.3ex\hbox{$>$\kern-.75em\lower1ex\hbox{$\sim$}}}}
\def\ltwid{\mathrel{\raise.3ex\hbox{$<$\kern-.75em\lower1ex\hbox{$\sim$}}}}
\def\square{\kern1pt\vbox{\hrule height 1.2pt\hbox{\vrule width 1.2pt\hskip 3pt
   \vbox{\vskip 6pt}\hskip 3pt\vrule width 0.6pt}\hrule height 0.6pt}\kern1pt}
\def\tdot#1{\mathord{\mathop{#1}\limits^{\kern2pt\ldots}}}

\def\pmb#1{\setbox0=\hbox{#1}%
  \kern-.025em\copy0\kern-\wd0
  \kern  .05em\copy0\kern-\wd0
  \kern-.025em\raise.0433em\box0 }

\catcode`@=11
\newcount\tagnumber\tagnumber=0

\immediate\newwrite\eqnfile
\newif\if@qnfile\@qnfilefalse
\def\write@qn#1{}
\def\writenew@qn#1{}
\def\w@rnwrite#1{\write@qn{#1}\message{#1}}
\def\@rrwrite#1{\write@qn{#1}\errmessage{#1}}

\def\taghead#1{\gdef\t@ghead{#1}\global\tagnumber=0}
\def\t@ghead{}

\expandafter\def\csname @qnnum-3\endcsname
  {{\t@ghead\advance\tagnumber by -3\relax\number\tagnumber}}
\expandafter\def\csname @qnnum-2\endcsname
  {{\t@ghead\advance\tagnumber by -2\relax\number\tagnumber}}
\expandafter\def\csname @qnnum-1\endcsname
  {{\t@ghead\advance\tagnumber by -1\relax\number\tagnumber}}
\expandafter\def\csname @qnnum0\endcsname
  {\t@ghead\number\tagnumber}
\expandafter\def\csname @qnnum+1\endcsname
  {{\t@ghead\advance\tagnumber by 1\relax\number\tagnumber}}
\expandafter\def\csname @qnnum+2\endcsname
  {{\t@ghead\advance\tagnumber by 2\relax\number\tagnumber}}
\expandafter\def\csname @qnnum+3\endcsname
  {{\t@ghead\advance\tagnumber by 3\relax\number\tagnumber}}

\def\equationfile{%
  \@qnfiletrue\immediate\openout\eqnfile=\jobname.eqn%
  \def\write@qn##1{\if@qnfile\immediate\write\eqnfile{##1}\fi}
  \def\writenew@qn##1{\if@qnfile\immediate\write\eqnfile
    {\noexpand\tag{##1} = (\t@ghead\number\tagnumber)}\fi}
}

\def\callall#1{\xdef#1##1{#1{\noexpand\call{##1}}}}
\def\call#1{\each@rg\callr@nge{#1}}

\def\each@rg#1#2{{\let\thecsname=#1\expandafter\first@rg#2,\end,}}
\def\first@rg#1,{\thecsname{#1}\apply@rg}
\def\apply@rg#1,{\ifx\end#1\let\next=\relax%
\else,\thecsname{#1}\let\next=\apply@rg\fi\next}

\def\callr@nge#1{\calldor@nge#1-\end-}
\def\callr@ngeat#1\end-{#1}
\def\calldor@nge#1-#2-{\ifx\end#2\@qneatspace#1 %
  \else\calll@@p{#1}{#2}\callr@ngeat\fi}
\def\calll@@p#1#2{\ifnum#1>#2{\@rrwrite{Equation range #1-#2\space is bad.}
\errhelp{If you call a series of equations by the notation M-N, then M and
N must be integers, and N must be greater than or equal to M.}}\else%
 {\count0=#1\count1=#2\advance\count1
by1\relax\expandafter\@qncall\the\count0,%
  \loop\advance\count0 by1\relax%
    \ifnum\count0<\count1,\expandafter\@qncall\the\count0,%
  \repeat}\fi}

\def\@qneatspace#1#2 {\@qncall#1#2,}
\def\@qncall#1,{\ifunc@lled{#1}{\def\next{#1}\ifx\next\empty\else
  \w@rnwrite{Equation number \noexpand\(>>#1<<) has not been defined yet.}
  >>#1<<\fi}\else\csname @qnnum#1\endcsname\fi}

\let\eqnono=\eqno
\def\eqno(#1){\tag#1}
\def\tag#1$${\eqnono(\displayt@g#1 )$$}

\def\aligntag#1\endaligntag
  $${\gdef\tag##1\\{&(##1 )\cr}\eqalignno{#1\\}$$
  \gdef\tag##1$${\eqnono(\displayt@g##1 )$$}}

\def\eqalignno#1{\displ@y \tabskip\centering
  \halign to\displaywidth{\hfil$\displaystyle{##}$\tabskip\z@skip
    &$\displaystyle{{}##}$\hfil\tabskip\centering
    &\llap{$\displayt@gpar##$}\tabskip\z@skip\crcr
    #1\crcr}}

\def\displayt@gpar(#1){(\displayt@g#1 )}

\def\displayt@g#1 {\rm\ifunc@lled{#1}\global\advance\tagnumber by1
        {\def\next{#1}\ifx\next\empty\else\expandafter
        \xdef\csname @qnnum#1\endcsname{\t@ghead\number\tagnumber}\fi}%
  \writenew@qn{#1}\t@ghead\number\tagnumber\else
        {\edef\next{\t@ghead\number\tagnumber}%
        \expandafter\ifx\csname @qnnum#1\endcsname\next\else
        \w@rnwrite{Equation \noexpand\tag{#1} is a duplicate number.}\fi}%
  \csname @qnnum#1\endcsname\fi}

\def\ifunc@lled#1{\expandafter\ifx\csname @qnnum#1\endcsname\relax}

\let\@qnend=\end\gdef\end{\if@qnfile
\immediate\write16{Equation numbers written on []\jobname.EQN.}\fi\@qnend}

\catcode`@=12

\catcode`@=11
\newcount\r@fcount \r@fcount=0
\newcount\r@fcurr
\immediate\newwrite\reffile
\newif\ifr@ffile\r@ffilefalse
\def\w@rnwrite#1{\ifr@ffile\immediate\write\reffile{#1}\fi\message{#1}}

\def\writer@f#1>>{}
\def\referencefile{
  \r@ffiletrue\immediate\openout\reffile=\jobname.ref%
  \def\writer@f##1>>{\ifr@ffile\immediate\write\reffile%
    {\noexpand\refis{##1} = \csname r@fnum##1\endcsname = %
     \expandafter\expandafter\expandafter\strip@t\expandafter%
     \meaning\csname r@ftext\csname r@fnum##1\endcsname\endcsname}\fi}%
  \def\strip@t##1>>{}}

\def\citeall#1{\xdef#1##1{#1{\noexpand\cite{##1}}}}
\def\cite#1{\each@rg\citer@nge{#1}}     

\def\each@rg#1#2{{\let\thecsname=#1\expandafter\first@rg#2,\end,}}
\def\first@rg#1,{\thecsname{#1}\apply@rg}       
\def\apply@rg#1,{\ifx\end#1\let\next=\relax
\else,\thecsname{#1}\let\next=\apply@rg\fi\next}

\def\citer@nge#1{\citedor@nge#1-\end-}  
\def\citer@ngeat#1\end-{#1}
\def\citedor@nge#1-#2-{\ifx\end#2\r@featspace#1 
  \else\citel@@p{#1}{#2}\citer@ngeat\fi}        
\def\citel@@p#1#2{\ifnum#1>#2{\errmessage{Reference range #1-#2\space is bad.}
    \errhelp{If you cite a series of references by the notation M-N, then M and
    N must be integers, and N must be greater than or equal to M.}}\else%
 {\count0=#1\count1=#2\advance\count1
by1\relax\expandafter\r@fcite\the\count0,%
  \loop\advance\count0 by1\relax
    \ifnum\count0<\count1,\expandafter\r@fcite\the\count0,%
  \repeat}\fi}

\def\r@featspace#1#2 {\r@fcite#1#2,}    
\def\r@fcite#1,{\ifuncit@d{#1}          
    \expandafter\gdef\csname r@ftext\number\r@fcount\endcsname%
    {\message{Reference #1 to be supplied.}\writer@f#1>>#1 to be supplied.\par
     }\fi%
  \csname r@fnum#1\endcsname}

\def\ifuncit@d#1{\expandafter\ifx\csname r@fnum#1\endcsname\relax%
\global\advance\r@fcount by1%
\expandafter\xdef\csname r@fnum#1\endcsname{\number\r@fcount}}

\let\r@fis=\refis                       
\def\refis#1#2#3\par{\ifuncit@d{#1}
    \w@rnwrite{Reference #1=\number\r@fcount\space is not cited up to now.}\fi%
  \expandafter\gdef\csname r@ftext\csname r@fnum#1\endcsname\endcsname%
  {\writer@f#1>>#2#3\par}}

\def\r@ferr{\endreferences\errmessage{I was expecting to see
\noexpand\endreferences before now;  I have inserted it here.}}
\let\r@ferences=\references
\def\references{\r@ferences\def\endmode{\r@ferr\par\endgroup}}

\let\endr@ferences=\endreferences
\def\endreferences{\r@fcurr=0
  {\loop\ifnum\r@fcurr<\r@fcount
    \advance\r@fcurr by 1\relax\expandafter\r@fis\expandafter{\number\r@fcurr}%
    \csname r@ftext\number\r@fcurr\endcsname%
  \repeat}\gdef\r@ferr{}\endr@ferences}


\let\r@fend=\endpaper\gdef\endpaper{\ifr@ffile
\immediate\write16{Cross References written on []\jobname.REF.}\fi\r@fend}

\catcode`@=12

\citeall\refto          
\citeall\ref            %
\citeall\Ref            %

\def\myspace{\baselineskip=\normalbaselineskip
  \multiply\baselineskip by 9 \divide\baselineskip by 5}

\def\ssc{\scriptscriptstyle}

\def\ufgrant{This work was supported in part by the
Institute for Fundamental Theory and by DOE contract
DE-FG05-86-ER40272.}

\def\dslash{\not{\hbox{\kern-2pt $\partial$}}}
\def\Dslash{\not{\hbox{\kern-4pt $D$}}}
\def\pslash{\not{\hbox{\kern-2.3pt $p$}}}

\phantom{
\cite{Nambu}
\cite{MTY}
\cite{Marciano}
\cite{BHL}
\cite{NJL}
\cite{topcolor}
\cite{LR}
\cite{CR}
\cite{KMP}
\cite{Bonisch}
\cite{King}
\cite{me}
\cite{CLB}
\cite{Luty}
\cite{Suzuki}
\cite{KM}
\cite{KingM}
\cite{Babu}
\cite{Barrios}
\cite{CCWBS}
\cite{BKMSWS}
\cite{HLP}
\cite{AD}
\cite{me0}
\cite{HHJKS}
\cite{LL}
\cite{tumbling}
\cite{chiralcolor}
\cite{tie1}
\cite{tie2}
\cite{Peskin}
\cite{quix}
\cite{quixphe}
\cite{custodial}
}
%
\oneandfourfifthsspace
\preprintno{UFIFT-HEP-92-7}
\preprintno{March, 1992}
\newstyletitle{A Tumbling Top-Quark Condensate Model}
\bigskip
\author Stephen P. Martin
\affil\uof

\body

\abstract
We propose a renormalizable
model with no fundamental scalars which breaks itself
in the manner of a ``tumbling" gauge theory
down to the standard model with a top-quark condensate. Because of
anomaly cancellation requirements, this model contains two
color sextet fermions (quixes), which are
vector-like with respect to the standard
model gauge group. The model also has a large number of pseudo-Nambu-Goldstone
bosons, some of which can be light.
The top-quark condensate is responsible for
breaking the electroweak gauge symmetry and gives the top quark a large mass.
We discuss the qualitative features and instructive shortcomings of the model
in its present form. We also show that this model can be naturally
embedded into an aesthetically pleasing
model in which the standard model fermion families appear symmetrically.

\endtitlepage
\myspace

Recently there has been a great deal of interest in the
idea[\cite{Nambu}-\cite{BHL}]
that the electroweak symmetry of the standard model is broken by
a top-quark condensate.
This would give a natural explanation
for the fact that the top quark has a much larger mass than any of the
other quarks and leptons, while simultaneously providing an
electroweak symmetry breaking
mechanism without a fundamental Higgs scalar.

In early versions of this idea, the top-quark
condensate was supposed to be induced
by a gauge-invariant but non-renormalizable four-fermion interaction
$$
{\cal L}_{\rm eff} \sim
{g^2\over M^2} \, ({\overline Q}^i_{\ssc } t_{\ssc })  \,
({\overline t}_{\ssc } Q_{{\ssc }\, i})
\eqno(4f)
$$
introduced at a scale $M$ which must be larger than the electroweak
breaking scale. [Here
$Q^i_{\ssc }  $ is the left-handed third generation
quark doublet, the $i$ is an $SU(2)_{\ssc L}$ index, and $t_{\ssc }$ is the
right-handed part of the top quark.
Color indices are suppressed.] If the coupling $g$ is  large enough
at the scale $M$, then a Nambu-Jona-Lasinio (NJL) mechanism[\cite{NJL}]
will trigger the
formation of a top-quark vacuum expectation
value
(``condensate")
$$
\langle {\overline Q}^i_{\ssc } t_{\ssc } \rangle =  \mu^3 \delta^{i1}
\eqno(qit)
$$
which breaks the electroweak symmetry.
The Higgs scalar boson is a composite top-anti-top
state bound by the interaction \(4f). The top-quark has a
Yukawa coupling to the composite Higgs which is of order
unity, and thus the top quark obtains a large mass.

The original formulation of the top-quark condensate idea is rather
unsatisfying
because it involves the {\it ad hoc} and non-renormalizable interaction
\(4f).
Now, there is an obvious precedent for four-fermion interactions
in elementary particle physics. The weak interactions were originally
described by an effective four-fermion interaction which was later
found to follow from integrating out massive intermediate vector
gauge bosons. Similarly, one can imagine that \(4f) is the result
of integrating out some heavy vector gauge bosons.
In this scenario, the top quark is heavy
because it couples to
a new gauge interaction which is strongly coupled and spontaneously
broken  at a mass scale larger than the electroweak
symmetry breaking scale. The most obvious way that this can happen is for the
new interaction to be an asymptotically free  non-abelian gauge theory.
Then the running gauge coupling constant will increase as we go to lower
mass scales. Eventually, the gauge coupling becomes large
enough to drive the formation of condensates, and the new gauge symmetry
is then spontaneously broken (e.g. by a mechanism to be proposed below)
so that it does not confine.
Several authors [\cite{topcolor}-\cite{me}] have
recently enumerated some possibilities
for the form of the renormalizable theory.
Other interesting extensions of and observations on the top-quark condensate
idea are found in [\cite{CLB}-\cite{LL}].

There are several important constraints on the
top-quark condensate scenario which come from demanding that it arise from a
renormalizable Lagrangian featuring a new non-abelian gauge interaction.
These follow from the simple observation
that if the top-quark has a special new gauge interaction, then other
fermions must also have that gauge interaction in order for the
full theory to be free of all gauge anomalies.
Generally, these fermions will be
``exotic", that is, they have transformation properties under the standard
model
gauge group which are different from the known quarks and leptons.
Of course, the prediction of new exotic fermions from the top-quark condensate
idea may be interesting if they are sufficiently heavy to have
avoided discovery until now, but not heavy enough to avoid discovery forever.
This can happen if the exotic fermions are in a complex representation of the
full gauge group including the new strongly coupled interaction, but transform
under the standard model subgroup as a real representation, so that they are
eligible to receive masses.
Note that one danger to be avoided in top-quark condensate model-building is
that
{\it a priori} these fermions might also participate in condensates which
could break the standard model gauge group in unacceptable ways.

The new strongly coupled gauge interaction will have an
approximate chiral symmetry which is spontaneously broken and includes the
electroweak symmetry as a subgroup. Then, as in technicolor models, there
will be a number of potentially light pseudo-Nambu-Goldstone bosons (PNGBs)
which are bound states of the fermions which couple to the strong gauge
interaction. These also may provide a means of experimental verification
or falsification of any particular model. The specific
properties of the extra fermions
and PNGBs of course depend on the particular model, but models
like the one we are going to consider here are always going to predict some
non-standard-model phenomena of this kind. Traditionally, the economy
of the top-quark condensate idea based on the non-renormalizable interaction
\(4f) has been used[\cite{BHL}] to make predictions involving constraints on
the
top-quark and Higgs masses. In contrast, the non-economy
implied by demanding renormalizability could provide a different kind
of prediction involving the existence of non-standard model particles.

To build a renormalizable top-quark condensate model, one may select a gauge
group $G$ which contains as a subgroup
the standard model gauge group $G_{\ssc SM} =
SU(3)_{c} \times SU(2)_{\ssc L} \times U(1)_{\ssc Y}$. The fermions
transform as an anomaly-free, complex representation of
$G$. This representation contains the usual standard model quarks and leptons
transforming in the usual way under $G_{\ssc SM}$, as well as some ``extra"
fermions which transform as a real represention of $G_{\ssc SM}$.
A simple subgroup
$H$
of $G$ becomes strongly coupled in the infrared,
producing the top-quark condensate and
possibly other condensates involving the other fermions which couple
to $H$. Now, one must also have the spontaneous symmetry breaking
$G \rightarrow G_{\ssc SM}$. Thus we are in a curious position: having
explained the cause of electroweak symmetry breaking
$G_{\ssc SM} \rightarrow G_0 = SU(3)_c \times U(1)_{\ssc EM}$ by means
of a top-quark condensate,
we must now explain
the origin of the symmetry breaking $G \rightarrow G_{\ssc SM}$!
The purpose of this paper is to propose a renormalizable top-quark condensate
model in which the spontaneous symmetry breaking occurs naturally without
any fundamental scalar fields. This is accomplished by arranging that one of
the ``other" condensates breaks $G \rightarrow G_{\ssc SM}$ while
the top-quark condensate breaks the electroweak symmetry. In
other words, the theory with gauge group $G$ and an appropriate fermion
representation automatically will break  itself in the pattern
$G \rightarrow G_{\ssc SM} \rightarrow G_0 $,
exactly in the manner of ``tumbling gauge theories"[\cite{tumbling}].
Indeed, we will find it most convenient to employ the language and
dynamical assumptions
of [\cite{tumbling}] in order to get a qualitative understanding of our
model.

We choose as a gauge group $G = SU(3)_1 \times SU(3)_2 \times SU(2)_{\ssc L}
\times U(1)_{\ssc Y}$. The standard model color $SU(3)_c$
is the diagonal subgroup of
$SU(3)_1 \times SU(3)_2 $.
This is exactly the gauge structure used in Hill's recent
``Topcolor" model[\cite{topcolor}].\footnote{$^\dagger$}{
This gauge group was
also earlier employed in models[\cite{chiralcolor}] which
have nothing to do with the top-quark condensate idea. }
However, we choose a different set of fermion assignments
for three reasons. First, the spontaneous symmetry breaking
$G \rightarrow G_{\ssc SM}$ will be an automatic consequence of the
condensation pattern given our choice of fermion representations, whereas
[\cite{topcolor}] requires a fundamental scalar (or some unspecified
dynamical mechanism) in order to provide this breaking. Second, as discussed
in [\cite{me}], the fermion representations in [\cite{topcolor}] contain a
real representation of the {\it unbroken} gauge group $G$. This means that
there are allowed
bare mass terms  in the case of [\cite{topcolor}] (one of which involves the
right-handed part of the bottom quark) even before symmetry breaking. In
order for [\cite{topcolor}] to work, one must make the assumption that those
mass terms are prohibited by an ungauged global symmetry whose
{\it raison d'etre}
remains mysterious.
Third, we will show at the end of this paper that our choice of fermion
representations
allows a natural extension to an aesthetically pleasing model which
treats the three families of quarks and leptons in a symmetrical way.

We assign fermions to the following representations of
$G = SU(3)_1 \times SU(3)_2 \times SU(2)_{\ssc L}\times U(1)_{\ssc Y}$:
$$
\eqalign{
& q_1\cr
& f_1, f_2 \cr
& Q^i \cr
& t \cr
& q_2 \cr
}
\eqalign{
 \> \sim \> &          \cr
 \> \sim \> &          \cr
 \> \sim \> &          \cr
 \> \sim \> &          \cr
 \> \sim \> &          \cr
}
\eqalign{
         &({\bf {\overline 6}}, {\bf 1}, {\bf 1}, -1/3) \cr
2 \times &({\bf 3}, {\bf 3}, {\bf 1}, 1/3) \cr
         &({\bf 3}, {\bf 1}, {\bf 2}, 1/6) \cr
         &({\bf {\overline 3}}, {\bf 1}, {\bf 1}, -2/3) \cr
         &({\bf 1}, {\bf {\overline 6}}, {\bf 1}, -1/3) \cr
}
\qquad \qquad
\eqalign{
& Q_1^i, Q_2^i \cr
& c, u \cr
& d   \cr
& L_1^i, L_2^i, L_3^i\cr
& \tau, \mu, e \cr
}
\eqalign{
 \> \sim \> &          \cr
 \> \sim \> &          \cr
 \> \sim \> &          \cr
 \> \sim \> &          \cr
 \> \sim \> &          \cr
}
\eqalign{
2 \times &({\bf 1}, {\bf 3}, {\bf 2}, 1/6) \cr
2 \times &({\bf 1}, {\bf {\overline 3}}, {\bf 1}, -2/3) \cr
         &({\bf 1}, {\bf {\overline 3}}, {\bf 1}, 1/3) \cr
3 \times &({\bf 1}, {\bf 1}, {\bf 2}, -1/2) \cr
3 \times &({\bf 1}, {\bf 1}, {\bf 1}, 1) \cr
}
\eqno(assign)
$$
[The gauge transformation properties of fermions are always given in terms
of left-handed two-component Weyl fields in \(assign) and throughout the rest
of
this paper.]
It is easy to check that all of the gauge anomalies cancel with this
fermion content.

How do the standard model fermions fit into \(assign)? After the symmetry
breaking $SU(3)_1 \times SU(3)_2 \rightarrow SU(3)_c$,
a fermion which transformed under
$SU(3)_1$ as $\bf R_1$ and under $SU(3)_2$ as
$\bf R_2$ will transform under the diagonal $SU(3)_c$
as the direct product representation ${\bf R_1} \times {\bf R_2}$.
So $f_1$ and $f_2$ each transform under $G_{\ssc SM}$ as
$({\bf\overline 3}, {\bf 1}, 1/3) + ({\bf 6}, {\bf 1}, 1/3)$. The two
copies of
$({\bf\overline 3}, {\bf 1}, 1/3)$
in $f_1$ and $f_2$
are identified as the charge conjugates of
two of the right-handed down-type quarks of the
standard model. It is easy to see that \(assign) contains
three standard model fermion families transforming under $G_{\ssc SM}$ as
$$
3 \times \left [ ({\bf 3}, {\bf 2}, 1/6) + ({\bf\overline 3}, {\bf 1}, -2/3)
+ ({\bf\overline 3}, {\bf 1}, 1/3) + ({\bf 1}, {\bf 2}, -1/2) +
({\bf 1}, {\bf 1}, 1) \right ]
\eqno(3fam)
$$
along with two vector-like quixes
$$
2 \times \left [({\bf 6}, {\bf 1}, 1/3) + ({\bf \overline 6}, {\bf 1}, -1/3)
\right ] \qquad .
\eqno(2quix)
$$
Note that the quixes are in a real representation of the standard model
gauge group and are thus eligible to receive masses after the symmetry
breaking $G\rightarrow G_{\ssc SM}$.
Also note that fractional electric charges are confined in this model.

In order to understand the symmetry breaking and generation of masses in this
model, let us now suppose that all of the gauge couplings are weak
at some sufficiently high energy scale and consider what happens as we
move to lower energy scales. Note that $SU(3)_1$ and $SU(3)_2$ are both
asymptotically free; their $\beta$-functions are given to one loop order by
$$
\eqalignno{
\beta_{\ssc 1} &= \mu {d g_{\ssc 1} \over d \mu} =
-{19\over 48 \pi^2} g_{\ssc 1}^3
&(beta1) \cr
{\rm and} \qquad \>\> \beta_{\ssc 2} &= \mu {d g_{\ssc 2} \over d \mu} =
-{15\over 48 \pi^2} g_{\ssc 2}^3
\qquad .
&(beta2)
\cr }
$$
Therefore it is quite reasonable to assume that $SU(3)_1$ becomes strongly
coupled first in the infrared, while the other couplings remain small. Thus
$SU(3)_1$ plays the role of $H$ in this model.

In order to understand the pattern of fermion condensation in our model,
we may turn to the dynamical assumptions outlined in [\cite{tumbling}], which
we now briefly review. Consider a model which consists of an
asymptotically free gauge theory which couples to some fermions
but no scalars. The fermions may also have weakly coupled gauge interactions
whose effects may be treated perturbatively.
When the strong gauge coupling becomes sufficiently large in the infrared,
a scalar fermion bilinear condensate will form in an irreducible
representation of the gauge group.
Suppose that the fermions involved
in the condensate transform under the strongly coupled gauge group in the
irreducible representations $R_1$ and $R_2$, and the resulting
condensate transforms as $R_s$.
(We treat all the fermions here as left-handed two-component
Weyl fermions.)
Thus $R_s$ occurs in the direct sum decomposition of the direct
product $R_1 \times R_2 = R_s + \cdots$.
We need a way of deciding for which choices of $R_1$, $R_2$, and $R_s$
the condensate will occur. According to the single gauge boson exchange
approximation, the condensate appears in
the ``most attractive scalar channel" (MASC),
$R_1 \times R_2 \rightarrow R_s$,
for which $V=  C_1 + C_2 - C_s$ is largest. Here $C_1$, $C_2$, and $C_s$
are the quadratic Casimir invariants for the representations
$R_1$, $R_2$, and $R_s$, respectively. [For example, if the strongly coupled
interaction were a $U(1)$, and left-handed fermions had charges $q_1$ and
$q_2$, then $V \propto  q_1^2 + q_2^2 -(q_1 + q_2)^2  = - 2q_1 q_2$, so that
for a collection of charged fermions, the most attractive channel occurs when
the product of charges is most negative. Thus in a general
gauge theory the statement that $V$ should
be maximized is the generalization of the
familiar statement in electrodynamics that opposite charges attract.]
The fermions which participate in the condensate obtain masses at this
stage, as do the gauge bosons corresponding to those generators of the
gauge symmetry which are spontaneously broken by the condensate.
The remaining gauge bosons and fermions define the next stage of the tumbling.

In the case of our model, the strongly coupled $SU(3)_1$ has LH fermions
transforming as a $\bf \overline 6$, eight $\bf 3$'s, and one $\bf \overline
3$.
The most attractive channels for this fermion content, and their relative
strengths $V$, are as follows:
$$
\eqalign{
& {\underline {\rm Channel}} \cr
&{ \bf\overline  6} \times  {\bf 3} \rightarrow { \bf \overline 3} \cr
&{ \bf\overline  6} \times  {\bf\overline 6} \rightarrow { \bf 6}  \cr
&{ \bf 3} \times {\bf \overline 3} \rightarrow {\bf 1}             \cr
&{ \bf\overline  6} \times  {\bf\overline 3} \rightarrow { \bf 8}  \cr
&{ \bf 3} \times {\bf  3} \rightarrow {\bf \overline 3}            \cr
}
\qquad
\eqalign{
& {\underline V} \cr
& 10 \cr
& 10 \cr
& 8  \cr
& 5  \cr
& 4 \qquad . \cr
}
\eqno(channels)
$$
{}From \(channels) we see that
the most naive version of the tumbling hypothesis is ambiguous, since there
is a tie for the MASC between the channels
${ \bf\overline  6} \times  {\bf 3} \rightarrow { \bf \overline 3}$ and
${ \bf\overline  6} \times  {\bf\overline 6} \rightarrow { \bf 6}$.
We must decide which of these condensates actually forms in order to
proceed.

Fortunately, other authors[\cite{tie1}-\cite{tie2}]
have already worried about what happens when there
is such a tie for the MASC in a tumbling gauge theory.
According to their criteria, the winner in our model is the channel
${ \bf\overline  6} \times  {\bf 3} \rightarrow { \bf \overline
3}$. More specifically, according to the arguments of
[\cite{tie2}] and [\cite{Peskin}], the condensate forms according to
$$
\langle q_1^{(\alpha\beta)} f_{1\gamma a} \rangle =
M^3 \delta^{(\alpha}_\gamma \delta^{\beta )}_a
\qquad .
\eqno(con1)
$$
(We use Greek letter $\alpha, \beta \ldots$ and Latin letters $a,b\ldots$
for indices in the fundamental representations of $SU(3)_1$ and $SU(3)_2$
respectively.) The composite scalar field
$\Phi^\alpha_a  = q_1^{(\alpha\beta)} f_{1\beta a}$ transforms under
$G$ as $({\bf\overline 3}, {\bf 3}, {\bf 1}, 0)$ and obtains a VEV
$\langle \Phi^\alpha_a \rangle = 2 M^3 \delta^\alpha_a$.
This condensate breaks $SU(3)_1 \times SU(3)_2 \rightarrow SU(3)_c$ as
promised.
Equation \(con1) reflects not only the assumed preference of the strongly
coupled theory for the channel
${ \bf\overline  6} \times  {\bf 3} \rightarrow { \bf \overline
3}$, but also the solution to a vacuum alignment problem[\cite{Peskin}],
namely which of the eight $\bf 3$'s of $SU(3)_1$ will condense with $q_1$.
There is a simple heuristic reason why the condensate chooses
to leave $SU(3)_c$ unbroken as in \(con1);
this is because the fermions participating
in the condensate \(con1) transform as a $\bf 6$ and a $\bf \overline 6$
of $SU(3)_c$ and thus feel an additional attractive force which would not
be present if the $q_1$ chose to condense in such a way as to break
$SU(3)_c$. (Of course,
the choice of $f_1$ instead of $f_2$ in \(con1) is completely arbitrary.)

Of the sixteen gauge bosons associated with $SU(3)_1 \times SU(3)_2$,
eight remain massless after the symmetry breaking and are the gluons
of QCD. The other eight gauge bosons obtain a mass of order $g_{\ssc 1} M$
and also transform as an octet of $SU(3)_c$. If one integrates out these
heavy gauge bosons, one obtains[\cite{topcolor}] precisely the four-fermion
interaction \(4f) which was our original motivation (along with some weaker
four-fermion interactions).
If the coupling constants of $SU(3)_1$ and $SU(3)_2$ at the scale $M$
are $g_{\ssc 1}$ and $g_{\ssc 2}$, then it is easy to show that the QCD
coupling
constant at $M$ is given by
$g_c = {g_{\ssc 1} g_{\ssc 2} / \sqrt{g_{\ssc 1}^2 + g_{\ssc 2}^2}}$.
We assume that
$SU(3)_{\ssc 1}$ is strongly coupled at $M$ and $SU(3)_{\ssc 2}$ is not, so
that
$g_{\ssc 1} \gg g_{\ssc 2}$ and $g_c \approx g_{\ssc 2}$.

According to \(con1), all of the components of $q_1$ condense, along with
the part of $f_1$ which transforms as a $\bf 6$ of $SU(3)_c$. This quix
receives a mass and decouples from the tumbling. Another quix, consisting
of $q_2$ and the part of $f_2$ which transforms as a $\bf 6$ of $SU(3)_c$,
remains uncondensed and massless at this stage. The uncondensed parts of $f_1$
and $f_2$ which transform as $\bf \overline 3$'s of $SU(3)_c$
are the charge conjugates of
two of the  right-handed down-type quarks of the standard model and
remain massless at this stage.

The next most attractive scalar
channel in \(channels) (not including the channel
${ \bf\overline  6} \times  {\bf\overline 6} \rightarrow { \bf 6}$,
because the ${ \bf\overline  6}$ has already condensed) is the
${ \bf 3} \times  {\bf\overline 3} \rightarrow { \bf 1}$.
Since the strength of the attraction in this channel is only slightly
less than that of the MASC, we make the dynamical assumption that this
condensate is also triggered even though \(con1) breaks $SU(3)_1$.
In fact, this corresponds to the assumption in the NJL language that
the four-fermion interaction \(4f) is sufficiently attractive to produce
a top-quark condensate. Now there is again a vacuum alignment problem
since the $\bf \overline 3$ has a choice of $\bf 3$'s with which to condense.
Again, the condensate will choose to avoid breaking $SU(3)_c$, so that the
condensate is of the form
$$
\langle Q^i_\alpha t^{\beta} \rangle =
\mu^3 \delta^{i1} \delta^\beta_\alpha
\qquad .
\eqno(con2)
$$
This is just the top-quark condensate which was our original motivation,
with color indices restored.
Heuristically, the theory prefers \(con2) because $Q^i$ and $t$ transform
as a $\bf 3$ and $\bf\overline 3$ of $SU(3)_c$, and this provides an additional
attractive force which would not be present for any other condensate
which breaks $SU(3)_c$. Of course, the $\delta^{i1}$ is just an arbitrary
choice of orientation of weak isospin. The
condensate \(con2) breaks $SU(2)_{\ssc L} \times U(1)_{\ssc Y} \rightarrow
U(1)_{\ssc EM}$ with the composite field
$\phi^i =Q^i_\alpha t^\alpha$ playing the
role of the standard model
Higgs scalar boson. The top quark condenses and gets a mass, as do
the $W^{\pm}$ and $Z^0$ vector bosons.

As we move
further into the infrared, the next interesting thing that happens is that
the remaining light quix condenses, due primarily to the QCD force,
and so obtains a large constituent mass. This condensate has the form
$$
\langle q_2^{(ab)} f_{2\alpha c} \rangle =
m^3 \delta^{(a}_\alpha \delta^{b)}_c
\qquad .
\eqno(con3)
$$
This condensate can occur at a much higher energy scale than for the ordinary
quarks in QCD, because the quadratic Casimir invariant of the $\bf 6$ of
$SU(3)$ is $5/2$ times that of the $\bf 3$. The constituent mass of the
lighter quix could therefore be as high as a few hundred GeV.
(This fact has been exploited by Marciano[\cite{quix}] who suggested that
a quix condensate driven by QCD
could be responsible for electroweak symmetry breaking.
The quix in our model plays a quite different role, since it is an
$SU(2)_{\ssc L}$ singlet and our quix condensates do not break $G_{\ssc SM}$.)
The lighter quix can also get a mass which is a current mass from the
point of view of the standard model interactions, by
 integrating out the heavy octet of gauge bosons, and by mixing with the
heavier quix due to some additional interactions at higher energies.

A quix will be pair-produced at hadron colliders by gluon fusion. In
our model, each quix will decay by emitting a heavy color octet
vector boson, turning into one of the down-type quarks  which are
components of $f_1$ and $f_2$. The heavy vector boson will then decay
into a quark-anti-quark pair. Thus the experimental signature for
the quix should consist of a six-jet signal above the QCD background.
For a quix with mass in the hundred GeV range, such a signal is
difficult but not impossible to detect at the Tevatron, LHC,
or SSC[\cite{quixphe}].

Let us now consider the spectrum of PNGBs which arise from our model.
The $SU(3)_1$ interaction has an approximate chiral symmetry
$SU(3)_1 \times SU(8) \times U(1) \times U(1)$. (There would be three
$U(1)$'s, but one of them has an $SU(3)_1$ anomaly.)
The first condensate \(con1) breaks this down to $SU(3)_c \times
SU(2)_{\ssc L} \times U(1)_{\ssc Y} \times U(1) \times U(1)$. There
are therefore 59 PNGBs from this stage, of which 8 are eaten and give
mass to the $SU(3)_c$ octet of heavy gauge bosons.
The second condensate \(con2) further breaks the chiral symmetry down to
$SU(3)_c \times U(1)_{\ssc EM} \times U(1)
\times U(1)$. There are three would-be Nambu-Goldstone bosons
from this stage which
are eaten by the $W^{\pm}$ and $Z^0$ vector bosons.
Finally, the lighter quix condensate
\(con3) breaks an additional $U(1)$, so that the original chiral symmetry
is broken down to $SU(3)_c \times U(1)_{\ssc EM}
\times U(1)_{\ssc B}$. The baryon number $U(1)_{\ssc B}$ is an
exactly conserved,
non-anomalous
global symmetry of the $SU(3)_1 \times SU(3)_2$ interactions (but has the usual
$SU(2)_{\ssc L}$ anomaly of the standard model). So there
are 52 uneaten PNGBs. Of these, 24 transform as eight $\bf 3$'s
and 24 more transform as three $\bf 8$'s  of $SU(3)_c$. These colored
PNGBs obtain large masses as in technicolor models. However, the remaining
four axion-like neutral PNGBs may be dangerously light.

The model we have described here clearly cannot be complete as it stands.
Perhaps the most glaring evidence of this is that the leptons remain
massless and in fact are decoupled from the symmetry breaking sector.
One might imagine that the lighter quarks and leptons can be given realistic
masses
by adding in higher order interactions analogous to those in extended
technicolor models. Such interactions might
have the additional beneficial effect of contributing to the masses of
the neutral PNGBs mentioned in the previous paragraph. Of course, one
may also expect to encounter the same problems that occur in extended
technicolor. For example, the required additional interactions
may give rise to flavor changing neutral current interactions at an
unacceptable level.
The most obvious way to try to couple the leptons to the symmetry breaking
sector is to embed $SU(3)_1$  into a Pati-Salam
$SU(4)$ at some high energy scale. There are several inequivalent ways to
embed the fermion content \(assign) into the enlarged gauge group; so far
we have not found any particularly satisfying way to do it.

Another potential disaster for our model involves the parameter
$\rho = M_{\ssc W}^2/M_{\ssc Z}^2 cos^2 \theta_{\ssc W}$,
which is constrained experimentally
to be very near 1. The usual way of ensuring this in dynamical
electroweak symmetry breaking models, as in the standard model with
a fundamental Higgs, is to arrange for a ``custodial"
$SU(2)$ symmetry[\cite{custodial}] of both the Lagrangian and the vacuum,
under which the generators of $SU(2)_{\ssc L}$ transform as a triplet.
Our model has no such custodial $SU(2)$.
However, the
situation may not be completely hopeless;
consider for example the scenario of [\cite{BHL}] in which $M$ is taken
to be $\gg 246$ GeV.
The effective theory far below $M$ looks
like the standard model with a heavy top quark and a Higgs doublet,
so that if the top quark is arranged to not be too heavy,
the $\rho$-parameter could come out in the allowed range.
Now, the
renormalization group methods used in [\cite{BHL}] rely for their
validity on the assumption that the theory is already fine-tuned, so that
the scale of new physics is much larger than the electroweak scale.
Since the avoidance of fine-tuning is one of the main motivations
for investigating dynamical electroweak symmetry breaking in the first
place, we tend to favor the opposite possibility, namely that the symmetry
breaking $G\rightarrow G_{\ssc SM}$ occurs at a scale not too far
removed from the scale of electroweak symmetry breaking. In this case,
the arguments used in [\cite{BHL}] are not reliable and should not be
used to draw quantitative conclusions; in particular the prediction of
a very heavy top might be avoided. If the scale $M$ is
sufficiently close to 246 GeV, there will certainly be no
range of energy scales at which the interaction \(4f) alone
comes close
to accurately reflecting the strong coupling dynamics.
Furthermore, the effective theory will contain a much more complicated
spectrum of composite resonances than just a Higgs doublet. These resonances
are also bound states of the fermions which couple to $SU(3)_1$.
There should be,
for example, composite vector particles exactly
analogous to the techni-$\rho$ and
techni-$\omega$ of technicolor models. (This has been emphasized already in
[\cite{LL}].)
Perhaps some presently
mysterious feature of the strong coupling dynamics prefers
$\rho \approx 1$. Or perhaps the value of $M$ is small enough to invalidate the
quantitative conclusions of [\cite{BHL}] without invalidating the qualitative
conclusion that  the effective theory below $M$ consists of the standard
model with one Higgs doublet and other resonances and
interactions which violate the custodial
$SU(2)$ in a controlled way.
To analyze whether these (perhaps optimistic) possibilities can be realized
requires an improved
understanding of the rather murky dynamics of strongly coupled
spontaneously broken theories, especially since the condensates \(con1) and
\(con2) have close to the same strength in the single gauge boson
approximation.

The model we have presented here can be embedded into a
very symmetric-looking
model by introducing another gauged $SU(3)$.
Thus we now take the unbroken gauge group to be
$G^\prime = SU(3)_1 \times SU(3)_2 \times SU(3)_3 \times SU(2)_{\ssc L}
\times U(1)_{\ssc Y}$, and we take the fermions to transform as:
$$
\eqalign{
& q_1,q_2,q_3\cr
& f_1,f_2,f_3 \cr
& Q^i_3, Q^i_1, Q^i_2  \cr
& t,c,u \cr
& L^i_1, L^i_2, L^i_3 \cr
& \tau,\mu ,e \cr
}
\eqalign{
 \> \sim \> &          \cr
 \> \sim \> &          \cr
 \> \sim \> &          \cr
 \> \sim \> &          \cr
 \> \sim \> &          \cr
 \> \sim \> &          \cr
}
\eqalign{
 &({\bf {\overline 6}}, {\bf 1},{\bf 1}, {\bf 1}, -1/3) +
  ({\bf 1},{\bf {\overline 6}}, {\bf 1}, {\bf 1}, -1/3) +
  ({\bf 1}, {\bf 1},{\bf {\overline 6}},  {\bf 1}, -1/3) \cr
 &({\bf 3}, {\bf 3}, {\bf 1}, {\bf 1}, 1/3) +
  ({\bf 3}, {\bf 1}, {\bf 3}, {\bf 1}, 1/3) +
  ({\bf 1}, {\bf 3}, {\bf 3}, {\bf 1}, 1/3) \cr
 &({\bf 3}, {\bf 1}, {\bf 1}, {\bf 2}, 1/6) +
  ({\bf 1}, {\bf 3}, {\bf 1}, {\bf 2}, 1/6) +
  ({\bf 1}, {\bf 1}, {\bf 3}, {\bf 2}, 1/6)  \cr
 &({\bf {\overline 3}}, {\bf 1}, {\bf 1},{\bf 1}, -2/3)+
  ({\bf {\bf 1},{\overline 3}}, {\bf 1}, {\bf 1}, -2/3)+
  ({\bf {\bf 1},{\bf 1},{\overline 3}}, {\bf 1}, -2/3)\cr
& 3 \times ({\bf 1},{\bf 1}, {\bf 1}, {\bf 2}, -1/2) \cr
& 3 \times ({\bf 1}, {\bf 1},{\bf 1}, {\bf 1}, 1) \cr
}
\eqno(assign2)
$$
Note that the fermion content is now invariant under
interchange of the three $SU(3)$'s. Furthermore, the colored
fermions are arranged in irreducible representations which each occur
only once. By analogy with the ``Topcolor" of [\cite{topcolor}] and the
``Chiral Color" of [\cite{chiralcolor}], it is tempting to refer to this
enlarged model as ``Family Color", since the three $SU(3)$ interactions
in $G^\prime$ are associated with the three families.
In order to recover our previous model, we just assume that
$SU(3)_2 \times SU(3)_3$ breaks down to the diagonal $SU(3)$ subgroup,
which is identified with the $SU(3)_2$ of $G$. [It is suggestive that such
a breakdown would be caused by a condensation of the
$ ({\bf 1}, {\bf 1},{\bf {\overline 6}},  {\bf 1}, -{1/3}) $ with
the $  ({\bf 1}, {\bf 3}, {\bf 3}, {\bf 1}, {1/3})$
exactly analogous to the condensate \(con1), if $SU(3)_3$ gets strong at a
very high scale. However, we think it
prudent to refrain from extending our dynamical assumptions too far beyond
the realm of the standard model, since as we have already noted, we are
missing (at least)
some major ingredient which is responsible for generating lepton masses.]
The remaining
unbroken gauge group is then $G$, and the fermion content is precisely
that of eq.~\(assign) plus a quix which is vector-like with respect to the
gauge group $G$ and therefore presumeably gets a large mass at this
stage. We find it encouraging that the somewhat haphazard-looking fermion
content given in \(assign) actually can come from the more attractive
\(assign2). This is of course just one of the possible extensions of the
basic model with gauge group $G$ and fermion content \(assign).

In this paper, we have described a model for dynamical electroweak
symmetry breaking which borrows from the old idea of tumbling gauge theories
and the younger top-quark condensate idea. We have not attempted to draw
any precise
quantitative conclusions, being content with the qualitative observation
that the gauge symmetries are broken in the correct way and that the top
quark obtains a large mass. In any case, we need additional model-building
ideas in order to have a chance for a realistic mass spectrum for the lighter
quarks and leptons, and additional technical ideas in order to calculate
reliably without fine-tuning in the strongly coupled theory.
The model we have discussed here is an example of a dynamical electroweak
symmetry breaking scheme which is similar to the technicolor idea, with
the top quark playing the role of a techniquark, but differs
in that
the strongly coupled gauge theory is
broken instead of confining in the infrared.
Note that there are, qualitatively speaking, three possible fates for
an asymptotically free non-abelian gauge theory in the infrared. The first
possibility is that the theory can become spontaneously broken before it
has a chance to become strong; this is the fate of $SU(2)_{\ssc L}$ in the
standard model. The second possibility is that the theory can become
strongly coupled and confining without being broken; we understand this because
it is what happens
to QCD in the standard model. It is also what
is supposed to happen in technicolor theories.
The third possibility is that the theory can become strong enough to produce
condensates, but is then spontaneously broken so that it does not confine.
There is no standard model
example of this, but there is also no good reason why such a thing could not
happen between the electroweak scale and the Planck scale. Despite its
shortcomings in the present incarnation, we hope that our model illustrates
how this third possibility could be responsible for breaking the electroweak
symmetry.
\vskip .2cm

I am grateful to Jonathan Bagger,
Paul Griffin, Chris Hill, Pierre Ramond, Dave Robertson and Pierre Sikivie
for helpful comments.
\ufgrant

\references

\refis{Nambu} Y.~Nambu,
EFI Report No. 88-39, 1988 (unpublished);
in {\it New Trends in Strong
Coupling Gauge Theories}, 1988 International Workshop,
Nagoya, Japan, edited by M.~Bando, T.~Muta, and K.~Yamawaki),
(World Scientific, Singapore, 1989);
EFI Report No. 89-08, 1989 (unpublished).

\refis{chiralcolor} P.~H.~Frampton and S.~L.~Glashow,
\journal Phys. Lett., B190, 157, 1987;
\journal Phys. Rev. Lett., 58, 2168, 1987.

\refis{tie1} D.~Amati and M.~A.~Virasoro,
\journal Phys. Lett., B99, 225, 1981.

\refis{tie2} V.~P.~Gusynin, V.~A.~Miransky and Yu.~A.~Sitenko,
\journal Phys. Lett., B123, 407, 1983.

\refis{BHL} W.~A.~Bardeen, C.~T.~Hill, and M.~Lindner,
\journal Phys. Rev., D41, 1647, 1990.

\refis{NJL} Y.~Nambu and G.~Jona-Lasinio,
\journal Phys. Rev., 122, 345, 1961.

\refis{MTY} V.~Miransky, M.~Tanabashi, and K.~Yamawaki,
\journal Phys. Lett., B221, 177, 1989;
\journal Mod. Phys. Lett. A, 4, 1043, 1989.

\refis{CLB} T.~E.~Clark, S.~T.~Love, and W.~A.~Bardeen,
\journal Phys. Lett., B237, 235, 1990.

\refis{Marciano} W.~J.~Marciano,
\journal Phys. Rev. Lett., 62, 2793, 1989;
\journal Phys. Rev., D41, 219, 1990.

\refis{quix} W.~J.~Marciano,
\journal Phys. Rev., D21, 2425, 1980.

\refis{BKMSWS} M.~Bando, T.~Kugo, N.~Maekawa, N.~Sasakura, Y.~Watabiki,
and K.~Suehiro,
\journal Phys. Lett., B246, 466, 1990.

\refis{Luty} M.~A.~Luty,
\journal Phys. Rev., D41, 2893, 1990.

\refis{quixphe} R.~S.~Chivukula, M.~Golden, and E.~H.~Simmons,
\journal Phys. Lett., 257B, 403, 1991;
\journal Nucl. Phys., B363, 83, 1991.

\refis{AD} Y.~Achiman and A.~Davidson,
\journal Phys. Lett., B261, 431, 1991.

\refis{Peskin}  S.~Weinberg,
\journal Phys. Rev., D13, 974, 1976;
M.~Peskin,
\journal Nucl. Phys., B175, 197, 1980;
J.~Preskill,
\journal Nucl. Phys., B177, 21, 1981.

\refis{custodial}  P.~Sikivie, L.~Susskind, M.~Voloshin and V.~Zakharov,
\journal Nucl. Phys., B173, 189, 1980.

\refis{tumbling} S.~Raby, S.~Dimopolous, and L.~Susskind,
\journal Nucl. Phys., B169, 373, 1980.

\refis{CR} D.~E.~Clague and G.~Ross,
\journal Nucl. Phys., B364, 43, 1991.

\refis{Babu} K.~S.~Babu and R.~N.~Mohapatra,
\journal Phys. Rev Lett., 66, 556, 1991.

\refis{Barrios} F.~Barrios, U.~Mahanta,
\journal Phys. Rev., D43, 284, 1991.

\refis{HLP} C.~T.~Hill, M.~A.~Luty, and E.~A.~Paschos,
\journal Phys. Rev., D43, 3011, 1991.

\refis{me0} S.~P.~Martin,
\journal Phys. Rev., D44, 2892, 1991.

\refis{CCWBS} M.~Carena, T.~Clark, C.~Wagner, W.~A.~Bardeen, and K.~Sasaki,
``Dynamical Symmetry Breaking and the Top-Quark Mass in the Minimal
Supersymmetric Standard Model", preprint FERMILAB-PUB-91/96-T.

\refis{me} S.~P.~Martin, ``Renormalizable Top-Quark Condensate Models",
Florida preprint
UFIFT-HEP-91-24, October 1991, to appear in Physical Review D.

\refis{Bonisch} R.~Bonisch,
``Gauge Created Top Quark Condensate and Heavy Top", Munich University preprint
LMU-91-03, August 1991.

\refis{King} S.~F.~King,
\journal Phys. Rev., D45, 990, 1992.

\refis{KingM} S.~F.~King and S.~H.~Mannan,
\journal Phys. Lett., B241, 249, 1990;
\journal Phys. Lett., B254, 197, 1991.

\refis{topcolor} C.~T.~Hill,
\journal Phys. Lett., B266, 419, 1991.

\refis{KMP} T.~K.~Kuo, U.~Mahanta and G.~T.~Park
\journal Phys. Lett., B248, 119, 1990.

\refis{HHJKS} A.~Hasenfratz, P.~Hasenfratz, K.~Jansen, J.~Kuti and Y.~Shen,
\journal Nucl. Phys., B365, 79, 1991.

\refis{LL} M.~Lindner and D.~L\"ust,
\journal Phys. Lett., B272, 91, 1991.

\refis{LR} M.~Lindner and D.~Ross,
\journal Nucl. Phys., B370, 30, 1992.

\refis{Suzuki} M.~Suzuki,
\journal Phys. Rev., D41, 3457, 1990;
\journal Mod. Phys. Lett., A5, 1205, 1990.

\refis{KM} P.~Kaus and S.~Meshkov,
\journal Phys. Rev., D42, 1863, 1990.

\endreferences

\endit\end